\documentclass{PoS}

\title{The POEMMA (Probe of Extreme Multi-Messenger Astrophysics) mission}

\ShortTitle{POEMMA}

\author{\speaker{Angela V. Olinto}$^1$\\
       Corresponding author e-mail: \email{aolinto@uchicago.edu}}
\author{J. H. Adams,$^2$ 
R. Aloisio,$^3$ 
L. A. Anchordoqui,$^4$ 
D. R. Bergman,$^5$ 
M. E. Bertaina,$^6$ 
P. Bertone,$^7$ 
F. Bisconti,$^8$
M. Bustamante,$^{9}$
M. Casolino,$^{10}$ 
M. J. Christl,$^7$
A. L. Cummings,$^3$  
I. De Mitri,$^3$
R. Diesing,$^1$
J. B. Eser,$^{11}$ 
F. Fenu,$^6$ 
C. Gu\'epin,$^{12}$ 
E. A. Hays,$^{13}$  
E. Judd,$^{14}$   
J. F. Krizmanic,$^{13}$ 
E. Kuznetsov,$^2$
A. Liberatore,$^6$ 
S. Mackovjak,$^{15}$   
J. McEnery,$^{13}$ 
J. W. Mitchell,$^{13}$  
A. Neronov,$^{16}$ 
F. Oikonomou,$^{17}$
A. N. Otte,$^{18}$ 
E. Parizot,$^{19}$ 
T. C. Paul,$^{4}$ 
J. S. Perkins,$^{13}$ 
G. Pr\'ev\^ot,$^{19}$ 
P. Reardon,$^{2}$ 
M. H. Reno,$^{20}$
M. Ricci,$^{21}$
F. Sarazin,$^{11}$ 
K. Shinozaki,$^{6}$ 
J. F. Soriano,$^4$
F. Stecker,$^{13}$ 
Y. Takizawa,$^{10}$
R. Ulrich,$^{22}$
M. Unger,$^{22}$
T. Venters,$^{13}$  
L. Wiencke,$^{11}$ 
R. M. Young$^{7}$\\
$^1$ The University of Chicago, Chicago, IL, USA;
$^2$University of Alabama, Huntsville, AL, USA; 
$^3$Gran Sasso Science Institute, L'Aquila, Italy; 
$^4$City University of New York, Lehman College, NY, USA; 
$^5$University of Utah, Salt Lake City, Utah, USA; 
$^6$Universit\'a di Torino, Torino, Italy; 
$^7$NASA Marshall Space Flight Center, Huntsville, AL, USA; 
$^8$ Istituto Nazionale di Fisica Nucleare, Sezione di Torino, Italy; 
$^{9}$Niels Bohr Institute, University of Copenhagen, DK-2100 Copenhagen, Denmark;
$^{10}$RIKEN, Wako, Japan; 
$^{11}$Colorado School of Mines, Golden, CO, USA; 
$^{12}$Institut dÕAstrophysique de Paris, Paris, France; 
$^{13}$NASA Goddard Space Flight Center, Greenbelt, MD, USA;  
$^{14}$Space Sciences Laboratory, University of California, Berkeley, CA, USA; 
$^{15}$Institute of Experimental Physics, SAS, Kosice, Slovakia;
$^{16}$University of Geneva, Geneva, Switzerland; 
$^{17}$European Southern Observatory, Garching bei M\"unchen, Germany;
$^{18}$Georgia Institute of Technology, Atlanta, GA, USA; 
$^{19}$APC-Universite de Paris 7, Paris, France; 
$^{20}$University of Iowa, Iowa City, IA, USA;
$^{21}$Istituto Nazionale di Fisica Nucleare - Laboratori Nazionali di Frascati, Frascati, Italy;
$^{22}$Karlsruhe Institute of Technology, Karlsruhe, Germany.\\}

\abstract{The Probe Of Extreme Multi-Messenger Astrophysics (POEMMA) is designed to observe  cosmic neutrinos (CNs) above 20 PeV and ultra-high energy cosmic rays (UHECRs) above 20 EeV over the full sky. The POEMMA mission calls for two identical satellites flying in loose formation, each comprised of a 4-meter wide field-of-view (45 degrees) Schmidt photometer. The hybrid focal surface includes a fast (1 ${\mu}$s)  ultraviolet camera for fluorescence observations and an ultrafast (10 ns) optical camera for  Cherenkov observations. 
POEMMA will provide new multi-messenger windows onto the most energetic  events in the universe, enabling the study of new astrophysics and  particle physics at these otherwise inaccessible energies.
}

\FullConference{36th International Cosmic Ray Conference -ICRC2019-\\
		July 24th - August 1st, 2019\\
		Madison, WI, U.S.A.}

\begin{document}

\section{POEMMA Extreme Multi-Messenger Science}
\begin{figure}
\begin{center}
\includegraphics [width=1\textwidth]{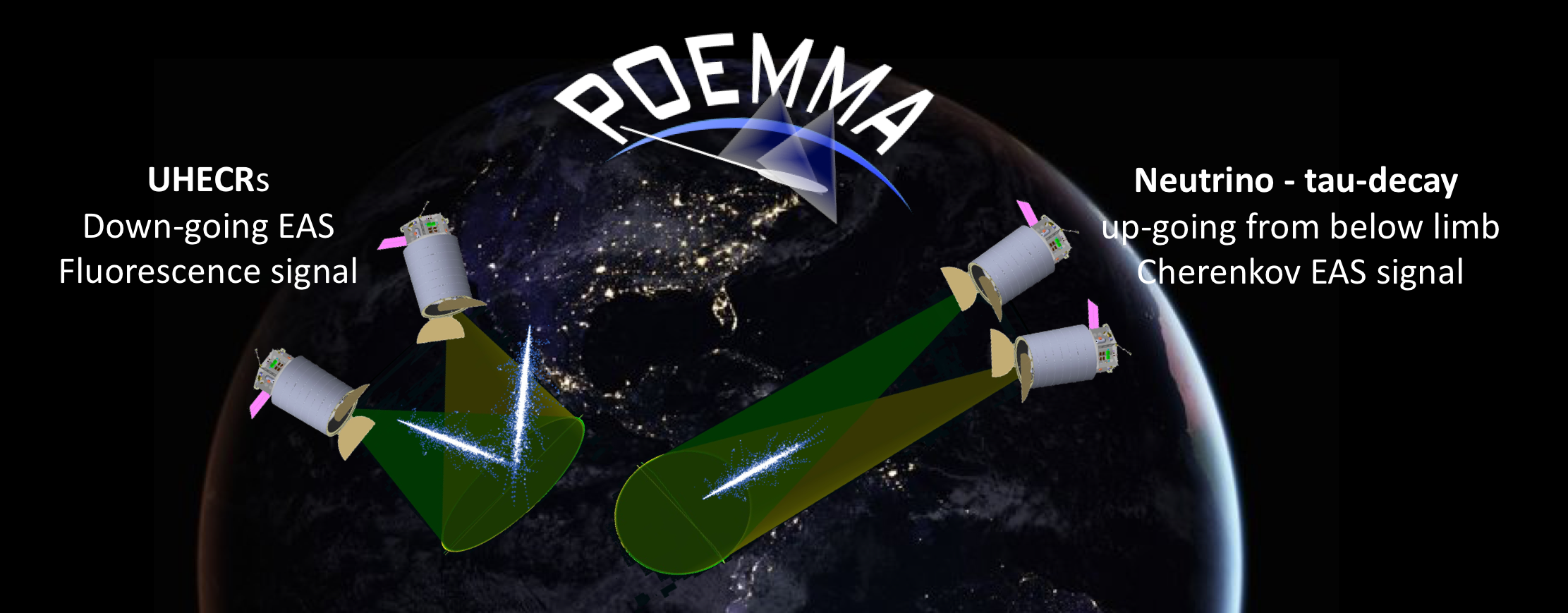}
\caption{POEMMA observing modes. Stereo fluorescence mode around the nadir on the left. On the right: Cherenkov mode from above (UHECRs) and below (cosmic tau neutrinos) the limb of the Earth.}
\label{fig2}
\end{center}
\end{figure}

The main scientific goals of  POEMMA  are to discover the elusive sources of cosmic rays with energies above 10$^{18}$ eV ($\equiv$ 1 EeV) and to observe cosmic neutrinos from multi-messenger transients with energies above 20 PeV. POEMMA exploits the tremendous gains in both ultra-high energy cosmic ray (UHECR) and cosmic neutrino exposures offered by space-based measurements, including the {\it full-sky coverage} of the celestial sphere.  For cosmic rays with energies $E \gtrsim 20~{\rm EeV}$, POEMMA will enable charged-particle astronomy by obtaining conclusive measurements of the UHECR spectrum,  composition, and source location. For multi-messenger transients, POEMMA will follow-up targets of opportunity transients to detect the first cosmic neutrino emission with energies $E_{\nu} \gtrsim$ 20 PeV. 
POEMMA also has sensitivity to neutrinos with energies above 20~EeV through fluorescence observations of neutrino induced EASs. Supplementary science capabilities of POEMMA include probes of physics beyond the Standard Model of particle physics, the measurement of $pp$ cross-section at 238 TeV center-of-mass energy, the study of atmospheric transient luminous events (TLEs), and the search for meteors and nuclearites. (See \cite{NASApoemma,Astro2020_poemma,UHECRpoemma,Venters:2019xwi,Reno2019icrc} for more details.)

\begin{figure}
\begin{center}
\includegraphics [width=1\textwidth]{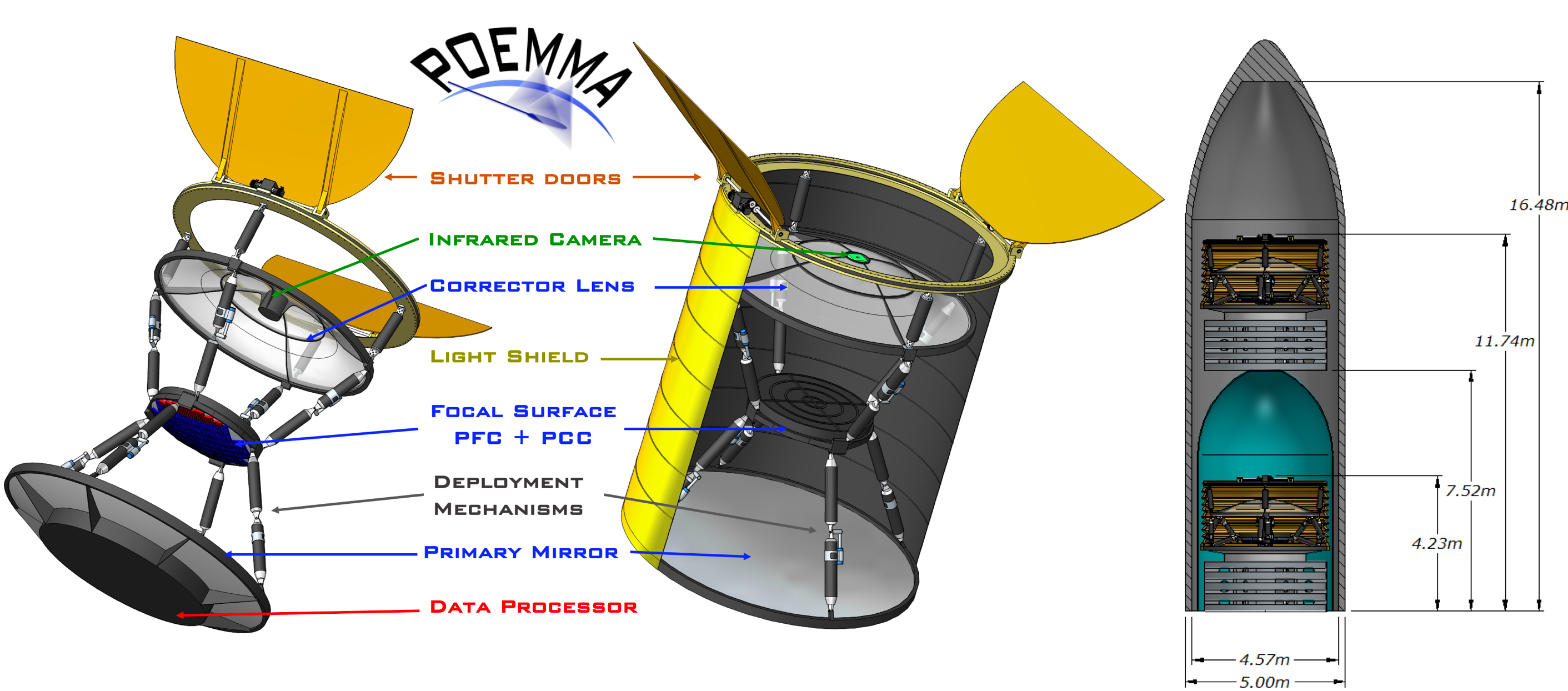}
\caption{Left: Concept of the POEMMA photometer with major components identified. Right: Both POEMMA photometers accommodated on Atlas V for launch. From~\cite{NASApoemma}.}
\label{fig1}
\end{center}
\end{figure}

The POEMMA can achieve this significant increase in sensitivity  by operating two observatories (described in Fig. \ref{fig1} and Table I) with very wide field of view in different orientation modes: a stereo fluorescence configuration close to the nadir for more precise UHECR observations and a tilted, Earth-limb viewing Cherenkov configuration for target of opportunity (ToO) neutrino searches (see Fig. \ref{fig2}). In limb observing mode, POEMMA can simultaneously search for neutrinos and UHECRs with Cherenkov observations, while observing UHECRs with fluorescence, thanks to the POEMMA focal surface hybrid design.  

POEMMA's fluorescence observations will yield one order of magnitude increase in yearly UHECR exposure compared to ground observatory arrays and two orders of magnitude compared to ground fluorescence telescopes. In the Cherenkov limb-viewing mode, POEMMA searches for optical Cherenkov signals of upward-moving EASs generated by $\tau$-lepton decays produced by $\nu_\tau$ interactions in the Earth with a terrestrial neutrino target of  $\sim 10^{10}$ gigatons. 
In the Cherenkov mode, an even more extensive  volume is monitored for UHECR fluorescence observations. Thus, {\bf POEMMA uses  the Earth and its atmosphere as a gargantuan high-energy physics  and astrophysics observatory.}

\begin{table}
\centering
\caption{POEMMA Specifications:}
\label{tab-1}  \
\begin{tabular}{lllllllll}
\hline
\hline
Photometer & Components &  &$\ \ $& Spacecraft  & \\ \hline
Optics &  Schmidt & 45$^\circ$ full FoV && Slew rate & 90$^\circ$  in 8 min \\
 & Primary Mirror & 4 m diam. && Pointing Res. & 0.1$^\circ$ \\
 & Corrector Lens & 3.3 m diam. && Pointing Know. & 0.01$^\circ$ \\
 & Focal Surface & 1.6 m diam. && Clock synch. & 10 ns \\  
 & Pixel Size & $3 \times 3$ mm$^2$  && Data Storage & 7 days \\ 
& Pixel FoV & 0.084$^\circ$ && Communication & S-band \\ 
PFC & MAPMT (1$\mu$s)& 126,720 pixels  && Wet Mass & 3,450 kg \\
PCC & SiPM (20 ns)& 15,360 pixels  && Power (w/cont)& 550 W \\ \hline
Photometer & (One)&  &&Mission  &(2 Observatories) \\ \hline
 & Mass & 1,550 kg  && Lifetime & 3 year  (5 year goal)\\
 & Power (w/cont) & 700 W   && Orbit & 525 km, 28.5$^\circ$ Inc \\
 & Data & $<$ 1 GB/day && Orbit Period & 95 min \\
& &  && Observatory Sep. & $\sim$25 - 1000 km \\\hline
\hline
\end{tabular}
 \
\center{Each Observatory = Photometer + Spacecraft; POEMMA Mission = 2 Observatories}
\end{table}

\subsection{UHECR Science}

\begin{figure}
\begin{center}
\includegraphics [width=1\textwidth]{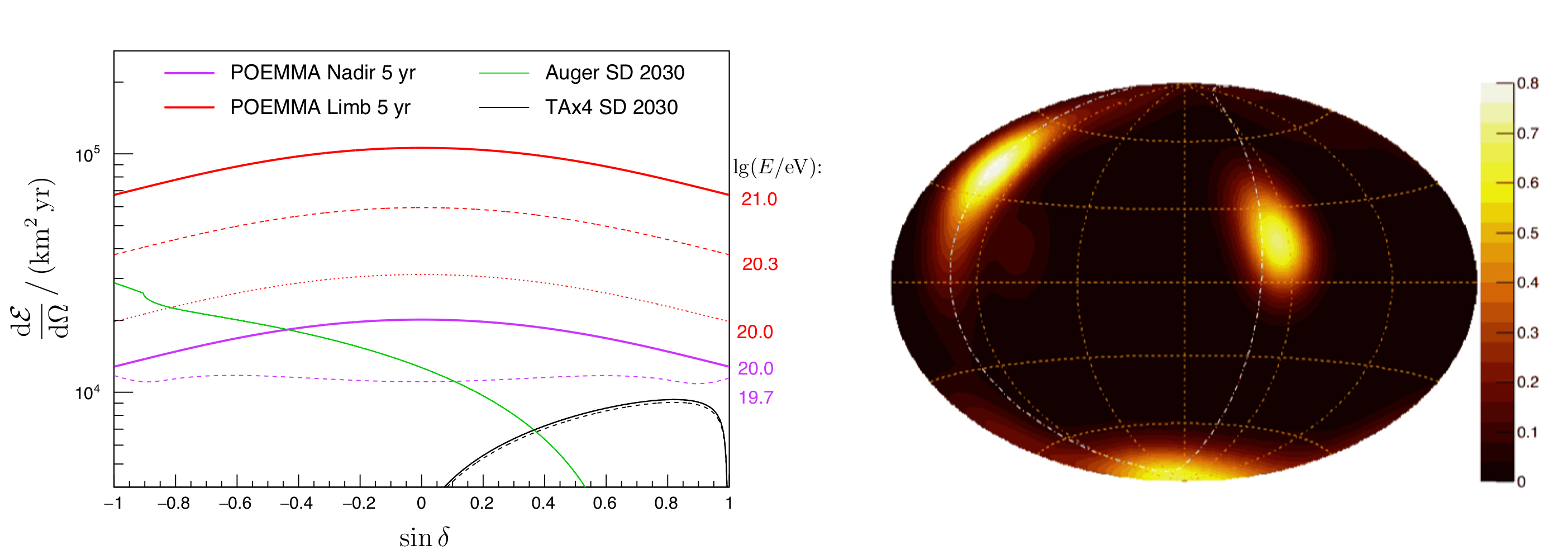}
\caption{Left: Differential exposure vs declination for POEMMA 5-yr in nadir  (purple) at 10$^{19.7}$ eV (dotted) and 10$^{20}$ eV (solid); and for limb (red) at 10$^{20}$ eV (dotted), 10$^{20.3}$ eV (dashed), and 10$^{21}$ eV (solid).  Exposures for Auger (green) and TAx4  (black) surface detectors (SD)  until 2030.
  Right: Probability density of the source sky map of nearby starburst galaxies with POEMMA exposure with $12.9^{\circ}$ smoothing. From~\cite{UHECRpoemma}.}
\label{figUHECR1}
\end{center}
\end{figure}

The nature of the astrophysical sources of UHECRs and their acceleration mechanism(s) remains a mystery as reviewed here~\cite{Kotera:2011cp,Anchordoqui:2018qom,Batista:2019}.  Proposed sources span a large range of astrophysical objects including extremely fast-spinning young pulsars, active galactic nuclei (AGN),
starburst galaxies (SBGs), and
gamma-ray bursts (GRBs). POEMMA data can determine which of the proposed model, if any, is the answer to this long-standing mystery.

 The two leading UHECR observatories currently in operation are the Pierre Auger
Observatory~\cite{Abraham:2010zz,Abraham:2009pm} in the southern
hemisphere, with $\sim 6.5 \times 10^4~{\rm km^2 \ sr \ yr}$ exposure over 13
years~\cite{Aab:2017njo}, and the Telescope Array (TA)~\cite{AbuZayyad:2012kk,Tokuno:2012mi} in the northern hemisphere, with $\sim 10^4~{\rm km}^2 \, {\rm sr} \, {\rm yr}$ exposure in 10 years (TA is currently upgrading by a factor of 4, named TAx4). As discussed in~\cite{UHECRpoemma}, 
POEMMA can reach  $\sim10^5$ to $ 10^6 ~{\rm km}^2 \, {\rm sr} \, {\rm yr}$ exposure in a 5 year mission, an increase of one to two (for fluorescence only)  orders of magnitude when compared with ground capabilities. POEMMA observes the full sky with the same instrument. Fig.~\ref{figUHECR1} left shows POEMMA's differential exposure as a function of declination compared to Auger and TAx4 extrapolated to 2030~\cite{UHECRpoemma}. 

\begin{figure}
\begin{center}
\includegraphics [width=1\textwidth]{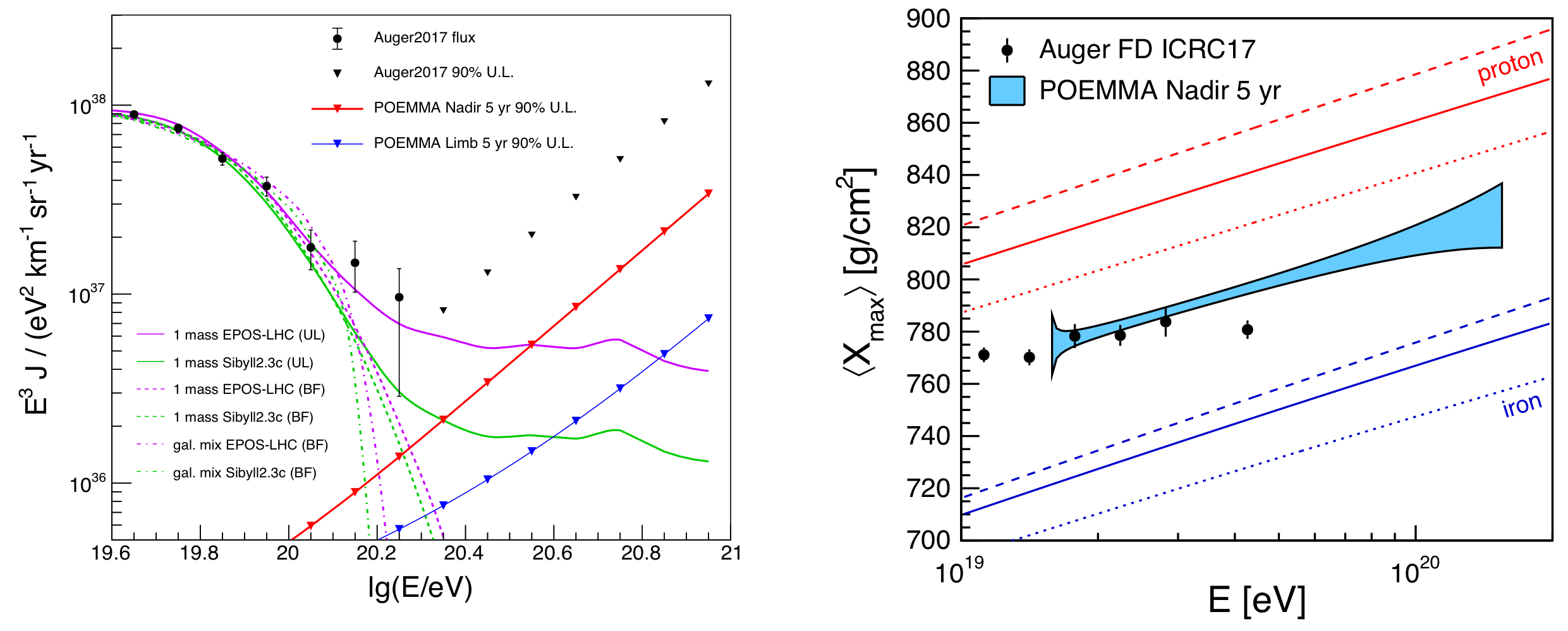}
\caption{Left: UHECR flux model predictions and Auger data~\cite{Aab:2017njo}. 90\% confidence upper limits are downward triangles: Auger 2017 (black), POEMMA 5 year stereo (red) and
  limb mode (blue).
     Right:  POEMMA capabilities for composition studies through measurements of the atmospheric column depth at which the EAS longitudinal development reaches maximum, $\langle X_{\rm max} \rangle$.  From~\cite{UHECRpoemma}.}
\label{figUHECR2}
\end{center}
\end{figure}

The powerful exposure provided by POEMMA is design to determine the sources of UHECRs  through the combined detailed observations of the sky distribution, the spectrum, and the composition at the highest energies. Fig.~\ref{figUHECR1} right shows an example of a POEMMA sky distribution of events if UHECR sources are located in nearby starburst galaxies.
POEMMA will  explore the differences in source models for the UHECR predicted spectrum above the current reach in energy, as shown in Fig.~\ref{figUHECR2} left.  In addition, POEMMA will measure the UHECR composition at 100s of EeV where models differ in predictions (Fig.~\ref{figUHECR2} right) further illuminating the origin of UHECRs.

\subsection{Cosmic Neutrino Science}

 In addition to observing the different components of UHECRs, the POEMMA design adds the novel capability of searching for neutrinos above 20 PeV (1 PeV $\equiv$ 10$^{15}$ eV)  from ToO events~\cite{Venters:2019xwi,Reno2019icrc,Guepin2018,Reno2019a}. (We use neutrinos to denote both neutrinos and anti-neutrinos, which interact similarly at these energies.)
POEMMA is unique in the ability to follow-up transient events on time scales of order one orbit (95 min) over the entire dark sky and on time scales of months over the full sky.

Very high-energy cosmic neutrinos are emitted in a number of models of astrophysical transient events~\cite{Meszaros:2017fcs,Ackermann:2019ows}. Astrophysical sources generally produce electron and muon neutrinos, which after astronomical propagation distances,  arrive on Earth with approximately equal numbers of the three flavors: electron, muon, and tau neutrinos. POEMMA detects primarily tau-neutrinos through the tau-decay generated EASs.  POEMMA will observe in general one third of the generated neutrino flux via the $\nu_\tau$ flux.
Since no prompt neutrinos have been observed at these energy scales, POEMMA will {\it discover} which transient events produce very-high energy neutrinos and at what times after the event. 

\begin{figure}
\begin{center}
\includegraphics [width=1\textwidth]{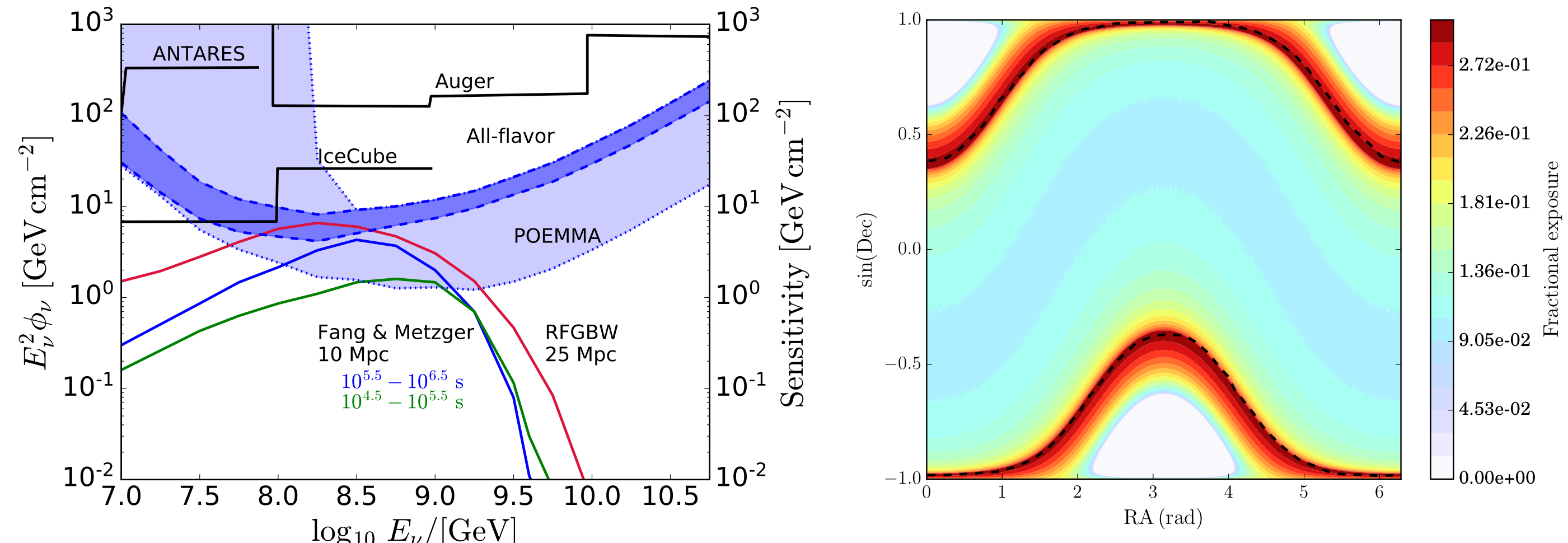}
\caption{Left: POEMMA ToO sensitivities to long bursts shown in purple. Dark
purple bands correspond to source locations between the dashed curves in the sky coverage figure on the right. Also shown IceCube, ANTARES, and Auger sensitivities (solid histograms), scaled to three flavors, for a 14 day time window around the binary neutron star merger GW170817~\cite{ANTARES:2017bia} and the  all-flavor fluence predictions for millisecond magnetars from Fang \& Metzger \cite{Fang:2017tla}. 
Right: POEMMA cosmic neutrino sky coverage, without including the $\sim 50\%$ blocking effect of the Sun~\cite{Guepin2018}, at a  given day of the year for viewing angles to $\delta = 18.3^\circ$ below the limb. Figures from~\cite{Venters:2019xwi}. }
\label{Neutr1}
\end{center}
\end{figure}

Examples of neutrino rich astrophysical transient events include short and long gamma-ray bursts, gravitational wave events from neutron-star binary coalescence, black hole-black hole coalescence (BH-BH),  the birth of pulsars and magnetars,  fast-luminous optical transients, blazar flares, tidal disruption events, and possibly many other high-energy transients. 
Neutrinos, not gamma rays, may be the primary emission signal in some environments (e.g., cosmic ray acceleration in white dwarf-white dwarf (WD-WD) mergers). 
POEMMA can follow up these events and reach a neutrino fluence around $E_{\nu}^2J_{\nu} \ge  0.1 \ {\rm GeV}/ {\rm cm}^{-2}$ depending on the location of the sources (see Fig.~\ref{Neutr1} and~\cite{Venters:2019xwi}). 

\section{POEMMA Instrument and Mission}

The POEMMA concept (Fig.~\ref{fig1}) evolved from previous work on the OWL~\cite{Stecker:2004wt} and the JEM-EUSO~\cite{JEM_EUSO} designs, the CHANT concept~\cite{Neronov:2016zou}, and the sub-orbital EUSO-SPB1 and 2~\cite{Wiencke:2017cfi,Adams:2017fjh}. 

The POEMMA instrument design~\cite{NASApoemma}  is comprised of two identical space-based platforms that detect extreme energy particles by recording the signals generated by EASs in the dark side of the Earth's atmosphere.  The central element of each POEMMA observatory  is a high sensitivity low resolution photometer that measures  two types of emission from these EASs: the faint isotropic emission due to the fluorescence of atmospheric nitrogen excited by air shower particles, and the brighter collimated Cherenkov emission from EASs directed at the POEMMA observatory.

\begin{figure}
\begin{center}
\includegraphics [width=1\textwidth]{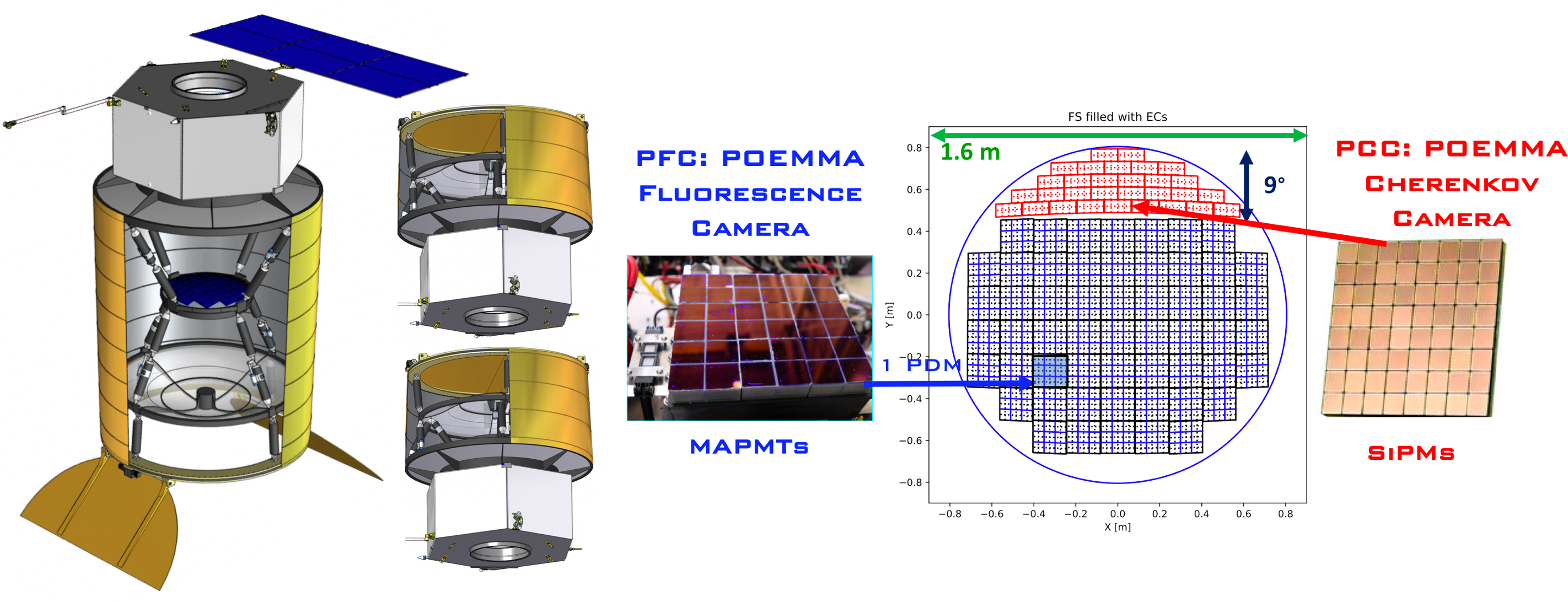}
\caption{Left: POEMMA instrument and spacecraft deployed and stowed for launch. Right: Layout of the photon sensors on the Focal Surface (PFC and PCC). From~\cite{NASApoemma}. }
\label{fig10}
\end{center}
\end{figure}

POEMMA  photometers are designed for deployment after launch. A stowed configuration (see Fig.~\ref{fig10}) enables two identical satellites to be launched together on a single Atlas V rocket (see Fig.~\ref{fig1}). Space qualified mechanisms extend each instrument after launch to their deployed position to begin observations.
The instrument architecture incorporates a large number of identical parallel sensor chains that meet the high standards of a Class B mission. Aerospace grade components have been identified for key elements within these sensor chains to insure reliability for the mission. 

\noindent{\bf Optics:} The POEMMA photometer  is based on a Schmidt optical design with a large spherical primary mirror (4 m diameter), the aperture and a thin refractive aspheric aberration corrector lens (3.3 m diameter) at its center of curvature, and a convex spherical focal surface (1.61 m diameter). This particular system provides a large collection aperture (6.39 m$^2$) and a massive field-of-view (45$^\circ$ full FoV).   The diameter of the POEMMA primary mirror is set to fit the launch vehicle (Atlas V). 
The PSF of the POEMMA optics is much less than a pixel size. POEMMA's imaging requirement is  $10^4$ away from the diffraction limit, implying optical tolerances closer to a microwave dish than an astronomical telescope. 

\noindent{\bf Focal Surface:}
The focal surface (FS) of the POEMMA photometer consists of the POEMMA Fluorescence Camera (PFC), optimized for the fluorescence signals, and the POEMMA Cherenkov Camera (PCC), optimized for for Cherenkov signals (see Fig.~\ref{fig10}).
The PFC records the EAS videos in 1$\mu$s frames in the $300 \lesssim \lambda/{\rm nm} \lesssim 500$ wavelength band using multi-anode photomultiplier tubes (MAPMTs).  Each MAPMT has 64  (3x3 mm$^2$) pixels in an 8 x 8 array.  The PFC is composed of 1,980 MAPMTs containing a total of 126,720 pixels. The stereo videos of the EAS determines the energy, direction, and composition of the UHECR.
The PCC  uses silicon photomultipliers (SiPMs) and is optimized to observe in the  $300 \lesssim \lambda/{\rm nm} \lesssim 900$ wavelength band for Cherenkov emission of showers developing towards the observatory. The PCC covers 9$^\circ$ of the FoV from the edge  (see Fig.~\ref{fig10}). PCC sensors are solid-state SiPMs  assembled in arrays of 8 x 8 pixels with a total area of 31 x 31 mm$^2$. The PCC records EASs produced by UHECRs above the limb of the Earth and showers from $\tau$-lepton decays below the Earth's limb induced by $\nu_\tau$ interactions in the Earth.

\noindent{\bf Spacecraft bus:}
The spacecraft bus  is a hexagonal cylinder 1.55 m tall with an outside diameter of 3.37 m weighing 1,073 kg. Located behind the POEMMA mirror, it provides basic services to the observatory including on-orbit deployment, power, communications, attitude control, propulsion and avionics.  The avionics includes the command and data handling system (including the flight computer), the spacecraft clock that provides the precise timing we need for synchronization between the satellites, the gimbal drive electronics to steer the solar panels, and the control functions for all the deployment mechanisms to unfold the instrument once it reaches orbit. 

\noindent{\bf Mission Concept:}
After calibration, the instruments will be pointed close to the nadir in order to make stereo observations of the fluorescent light from cosmic ray extensive air showers (EASs) at the lowest energies. Once sufficient statistics have been acquired at the lowest energies, the satellites separation will be reduced to $\sim$ 30 km and the instruments will be pointed for limb observations via Cherenkov. Throughout the mission instruments will be re-oriented towards neutrino ToO directions following a transient event alert. During a ToO follow-up mode, measurements of fluorescence from EASs will continue utilizing the larger volumes of the atmosphere observed from the satellites to the limb. 

\noindent{\bf Launch Operations:}
The POEMMA mission is designed for launch into circular orbits at an inclination of 28.5 degrees and an altitude of 525 km until de-orbited at the end of the mission. 
The satellites are launched in a stowed configuration. Once on orbit, the corrector plate and focal surface are deployed into their final position and the solar arrays are deployed from each spacecraft bus.

\noindent{\bf On-Orbit Operations:}
The satellites will orbit the Earth with a period of 95 minutes, or $\sim$15 times per day.  During observations the attitude of the satellites must be maintained within 0.1$^\circ$ with knowledge of the attitude to within 0.01$^\circ$. Events will be time-tagged with a relative accuracy within 25 ns between the satellites. 
The satellites have star trackers and sun sensors for accurate attitude knowledge.

\medskip

In sum, POEMMA is a unique probe-class mission designed to answer fundamental open questions in the multi-messenger domain starting in the next decade.

\medskip
\noindent{\bf Acknowledgements}
The POEMMA conceptual study was supported by NASA Grant NNX17AJ82G.

\end{document}